\documentclass[conference,comsoc]{IEEEtran}
\usepackage[T1]{fontenc}% optional T1 font encoding
\usepackage{amsmath}
\usepackage[cmintegrals]{newtxmath}
\usepackage{url}
\usepackage{soul,color,graphicx,float,epstopdf}
\usepackage{tablefootnote,textcomp,gensymb}
\usepackage{eurosym,booktabs,multirow}
\usepackage{tabularx} % Add this package

\usepackage{amsfonts} 
\usepackage{array}
\usepackage{stfloats}
\usepackage{float}
\usepackage{mathtools}
\usepackage{graphicx}
\usepackage{pdfpages}
\usepackage{comment}
\usepackage{balance}
\usepackage{xcolor}
\usepackage{comment,enumitem}
\usepackage[normalem]{ulem}
\usepackage{subfigure}
\definecolor{Orange}{rgb}{1,0.5,0}
\newcommand{\todo}[1]{\textsf{\textbf{\textcolor{Orange}{[#1]}}}}

\usepackage{algorithm}
\usepackage{algpseudocode}
\usepackage[implicit=false]{hyperref}

\begin{document}
\title{RL-Driven Semantic Compression Model Selection and Resource Allocation in Semantic Communication Systems}

\author{\IEEEauthorblockN{
Xinyi Lin, Peizheng Li, Adnan Aijaz
}\\ 
\vspace{-3.00mm}
\IEEEauthorblockA{
Bristol Research and Innovation Laboratory, Toshiba Europe Ltd., U.K.\\
Email: {\{xinyi.lin, peizheng.li, adnan.aijaz\}@toshiba-bril.com}}
\vspace{-8mm}}

\maketitle

\begin{abstract}
Semantic communication (SemCom) is an emerging paradigm that leverages semantic-level understanding to improve communication efficiency, particularly in resource-constrained scenarios. However, existing SemCom systems often overlook diverse computational and communication capabilities and requirements among different users. Motivated by the need to adaptively balance semantic accuracy, latency, and energy consumption, this paper presents a reinforcement learning (RL)-driven framework for semantic compression model (SCM) selection and resource allocation in multi-user SemCom systems. To address the challenges of balancing image reconstruction quality and communication performance, a system-level optimization metric called Rate-Distortion Efficiency (RDE) has been defined. The framework considers multiple SCMs with varying complexity and resource requirements. A proximal policy optimization (PPO)-based RL approach is developed to dynamically select SCMs and allocate bandwidth and power under non-convex constraints. Simulations demonstrate that the proposed method outperforms several baseline strategies. This paper also discusses the generalization ability, computational complexity, scalability, and practical implications of the framework for real-world SemCom systems.
\end{abstract}


\section{Introduction}
\label{sec:introduction}
Traditional communication systems struggle to meet diverse service demands under limited wireless resources, highlighting the need for a shift toward AI-enabled 6G networks. Semantic communication (SemCom) is expected to address this by transmitting the meaning of information rather than raw data, enabling goal-oriented, context-aware transmission that improves efficiency in noisy, bandwidth-constrained environments~\cite{10597087}. Applications of SemCom range from human-machine interaction~\cite{app12031192} and autonomous systems~\cite{10328182} to 6G networks~\cite{9937052} and Internet of Things (IoT) devices~\cite{10183798}, where understanding and interpreting the conveyed information is considered instead of transmitting merely the raw data. 

%With the constraints of limited wireless resources, traditional communication systems face the challenge of meeting the diverse service demands across various application scenarios, driving the need for a transformative shift in the era of AI-enabled 6G networks \cite{9797984}. Semantic communication (SemCom) is an advanced communication paradigm that goes beyond the traditional Shannon-based framework by focusing on conveying the meaning of the information. This ensures that the received message is not only accurate but also relevant and useful for the intended task \cite{10597087}. By incorporating goal-oriented and context-aware transmission, SemCom significantly enhances communication efficiency, particularly in bandwidth-limited and noisy environments. Applications of SemCom range from human-machine interaction \cite{app12031192} and autonomous systems \cite{10328182} to 6G networks \cite{9937052} and Internet of Things (IoT) devices \cite{10183798}, where understanding and interpreting the conveyed information is considered instead of transmitting merely the raw data.

Recent research has explored several promising topics in SemCom, including joint source-channel coding (JSCC)~\cite{10328187}, where deep learning (DL) techniques, such as transformers and variational autoencoders (VAE), are used to encode and decode semantic information robustly over noisy channels. %Semantic-aware edge computing is also gaining attention, leveraging edge devices to perform semantic extraction and processing closer to the data source, reducing latency and bandwidth consumption \cite{10024766}. 
% Additionally, semantic information theory is being developed to quantify the semantic content of messages, enabling more efficient communication systems \cite{niu2024mathematical}. 
SemCom for 6G networks is another cutting-edge topic, focusing on integrating SemCom with emerging technologies like AI, blockchain, and reconfigurable intelligent surfaces (RIS) to enable ultra-reliable, low-latency, and energy-efficient communication for applications such as autonomous vehicles, smart cities, and the metaverse~\cite{li2023open,10163877,10079087}. These advancements are driving the evolution of communication systems toward more intelligent, context-aware, and user-centric designs.

Although there are several preliminary research investigations on SemCom from an application-level perspective, challenges related to the sustainability of the overall SemCom system operation and the latency of the service delivery have not been fully explored yet. Existing works primarily focus on optimizing semantic extraction, reconstruction, and resource allocation for specific tasks, such as image or text transmission. For example, the semantic spectral efficiency (S-SE) was first defined in~\cite{9763856}, used to optimize resource allocation in terms of channel assignment and the number of transmitted semantic symbols. A framework where a set of servers cooperatively transmit image data to a set of users utilizing SemCom was investigated in~\cite{10149174}, which enables servers to transmit only the semantic information that accurately captures the meaning of images. To meet the quality of service (QoS) requirement of each user, physical layer-centric optimization, such as resource blocks (RBs), power and bandwidth used for semantic information transmission, was jointly optimized. In addition, the authors in~\cite{9450827} design a DL-enabled SemCom system based on an attention mechanism by utilizing a squeeze-and-excitation (SE) network to improve the recovery accuracy of speech signals, named DeepSC-S. However, these works often overlook the broader implications of energy consumption and computational efficiency in real-world deployments. As the adoption of AI-based semantic models grows, their energy-intensive nature raises significant concerns about the sustainability of large-scale systems, particularly in resource-constrained environments like edge networks. Additionally, the latency introduced by complex neural network (NN)-based semantic models and wireless transmission can hinder real-time service delivery, which is critical for time-sensitive applications like autonomous driving, healthcare, and virtual reality. To address these gaps, our work aims to bridge the divide between application-level performance and system-level sustainability by developing a framework that optimizes SemCom under the constraints of wireless resources, energy consumption and latency requirements.

In this paper, without the loss of generality, we consider a scalable SemCom system that involves multiple users, radio access network (RAN) and an edge server to process the image transmission task. The tasks of image compression and reconstruction are performed by NN-based models, referred to as semantic compression models (SCMs), which are designed to extract and reconstruct the semantic content of images. Potentially, diverse types of NN models can serve as a semantic encoder/decoder. The architectural complexity of these NN models determines the quality, speed and energy consumption of image compression/reconstruction. The optimization of these three factors, in conjunction with the radio recrudesces, is important for the service quality of the SemCom system, especially when it expands to a large scale.

The main contributions of this paper are summarized as follows.
\begin{itemize}
    \item To ensure QoS of image transmission, we define a metric \emph{rate-distortion efficiency} (RDE), which is used to evaluate the efficiency of a communication system in terms of balancing data transmission rate and distortion. It measures how well the system can transmit information while minimizing distortion in the received data. Our goal is to minimize RDE by optimizing the selection of SCMs, each with distinct compression rates and energy consumption, alongside power and bandwidth allocation.
    \item We consider various semantic models, specifically different convolutional neural network (CNN) architectures trained under a VAE framework, tailored for image-related tasks. The system must balance the trade-offs between AI model complexity, energy consumption, and reconstruction quality, as denser NN architectures increase inference time and energy usage, impacting both sustainability and service latency.
    \item Due to the dynamic nature of the environment, coupled discrete and continuous variables, and large scalability, we leverage proximal policy optimization (PPO)-based algorithm to optimize resource allocation and intend to achieve a balance between service quality, energy efficiency, and bandwidth consumption, ensuring reliable service delivery at acceptable energy and latency levels. %This approach addresses the challenges of large-scale AI model adoption, sustainability goals, and the need for efficient computational and wireless resource utilization in a wireless environment.
\end{itemize}

% \PL{The following sections of this paper are organized as follows:}

% \todo{To be added if necessary, let's do that at the end.}

\section{System Model}
\begin{figure}[t]
    \centering    \includegraphics[width=0.8\linewidth]{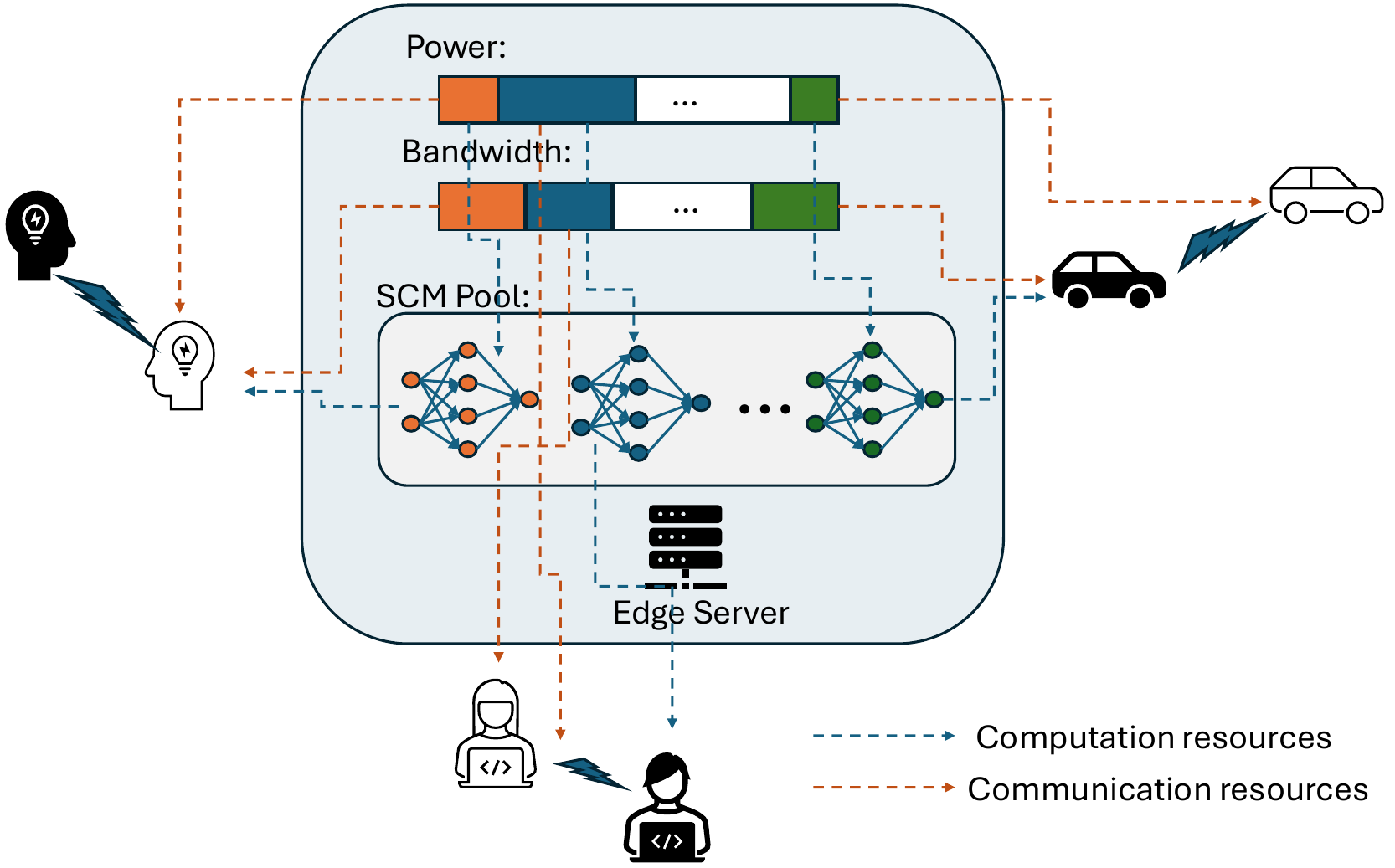}
    \caption{System model of a SemCom-powered wireless network with SCM selection and computation and communication resources allocation.}
    \label{systemmodel}
    \vspace{-5mm}
\end{figure}
As shown in Fig.~\ref{systemmodel}, we envisage a SemCom-powered cellular network comprising an edge server and a set of resource-constrained devices (users) $\mathcal{M}=\{1,...,M\}$. In this setup, we assume that the semantic encoding and decoding for all transmissions are performed on the devices. Each device holds a variety of SCMs $\mathcal{S}=\{1,\ldots,S\}$, each with different computational demands and compression performances tailored for image transmission. We use matrix $\boldsymbol{A}\in \mathcal{C}^{M\times S}$ to measure the SCM selection. Particularly, $a_{m,s}=1$ denotes that the $s$-th SCM is selected by user $m$, otherwise  $a_{m,s}=\ 0$. Each user can only select one SCM for a specific image transmission task. In addition, the power consumption and inference time of the $s$-th SCM are given by $p_{m,c}$ and $\tau_{m,c}$, respectively. The JSCC~\cite{10849728} scheme is applied, which aims to eliminate the traditional separation of source and channel coding, while optimizing them simultaneously in SemCom. During the end-to-end SemCom process, the sender first selects an appropriate SCM for encoding, and this process can be represented as
\begin{equation}
I_s^\prime=\mathcal{C}_{\theta_s}\left(I_s\right),
\end{equation}
where $\theta_s$ denotes the encoding parameter of the $s$-th SCM. $I_s^\prime$ denotes the encoded semantic features, which is further reshaped into $b$ complex-valued
symbols to form the encoded signal $x$. Then the encoded signal is transmitted over the wireless channels, i.e., $y = hx + n$, where the communication channels $h$ are integrated as non-trainable layers into the whole deep JSCC framework to enable end-to-end training.
On the receiver side, the receiver first
performs channel equalization to recover the transmitted signal and reshapes it to be $\widehat{{I_s}^\prime}$. Then, the same SCM is used to recover the semantic information from the received bits $\widehat{{I_s}^\prime}$ and reconstruct the images, which can be interpreted as
\begin{equation}
    \widehat{I_s}=\mathcal{D}_{\alpha_s}\left(\widehat{{I_s}^\prime}\right).
\end{equation}
The encoded source information with extracted semantic content is packed into binary bits for transmission over the wireless channel. Particularly, for the $m$-th user, the transmission rate can be expressed as
\begin{equation}
    r_m=\log_2{\left(1+\frac{p_{m,s}g_m}{B_mN_0}\right)},
\end{equation}
where $p_{m,s}$, $g_m$, $B_m$, and $N_0$ denote the transmission power, channel gain, allocated bandwidth, and the noise power, respectively. Based on the transmission rate, the transmission latency can be interpreted as
\begin{equation}   \tau_{m,t}=\frac{\left|x\right|}{r_m}.
\end{equation}
To ensure recovery quality, we introduce a unified image recovery distortion metric based on the cross-entropy loss function. Specifically, we model the recovery distortion of the image sent by the $m$-th user as
\begin{equation}
    D_m = \lVert I_s-\widehat{I_s}\rVert_2.
\end{equation}
As both the transmission rate and the distortion of images significantly affect the service quality, we define RDE, which intends to realize a trade-off between the two metrics, and the numerical expression of RDE is given as
\begin{equation}
    \text{RDE} = \frac{\sum_{m=1}^M r_m}{\lambda\sum_{m=1}^M D_m+\varepsilon},
\end{equation}
where $\lambda$ controls how much distortion is penalized in the efficiency calculation and $\varepsilon$ prevents numerical instability when the total distortion is small.

The optimization objective of the SCM selection and resource allocation problem is to maximize RDE. Moreover, we consider the bandwidth, power, and also latency constraints. Let $B_{max}$ and $P_{max}$ be the total bandwidth and energy in the considered cellular network. $\boldsymbol{B}\in\mathcal{C}^{1\times M}$ and $\boldsymbol{P}\in\mathcal{C}^{2\times M}$ denote the bandwidth and energy allocation metrics, respectively. For each user $m$, the maximum latency is denoted as $\tau_{m,max}$. The optimization problem can be described as
\begin{subequations}
\begin{align}
\begin{split}
\mathbb { P }: & \max\limits_{\boldsymbol{A},\boldsymbol{P},\boldsymbol{B}} \quad  \frac{\sum_{m=1}^M r_m}{\lambda\sum_{m=1}^M D_m+\varepsilon}
\end{split}\\
\begin{split}
 & \ \text { s.t. }  \ \quad\sum_{m=1}^M (p_{m,c}+p_{m,t})\le P_{max},\forall m\in\mathcal{M}, 
\end{split}\\
\begin{split}
 & \quad \ \qquad \sum_{m\in\mathcal{M}} B_m\le B_{max},
\end{split}\\
\begin{split}
 & \qquad \qquad \tau_{m,c}+\tau_{m,t}\le\tau_{m,max},\forall m\in\mathcal{M},
 \end{split}\\
 \begin{split}
 & \qquad \qquad a_{m,s}\le1,\forall s\in\mathcal{S},
 \end{split}\\
  \begin{split}
 & \quad \ \ \qquad \sum_{s\in\mathcal{S}} a_{m,s}=1,\forall m\in\mathcal{M}.
 \end{split}
\end{align}
\vspace{-3mm}
\end{subequations}

The considered optimization problem jointly addresses discrete SCM selection and continuous resource allocation under nonlinear constraints, making it highly non-convex and difficult to solve with traditional methods. PPO is well-suited for this setting as it can effectively learn policies in hybrid action spaces without requiring differentiable models, enabling stable and scalable optimization of long-term reward in complex semantic communication systems. In the following, we propose a PPO-based adaptive SCM selection and resource allocation strategy, which is robust to the dynamic environment and user requirements.

\section{PPO-Based SCM Selection and Resource Allocation}

\subsection{Training of SCM Models} 
\begin{figure}[t]
    \centering    \includegraphics[width=0.8\linewidth]{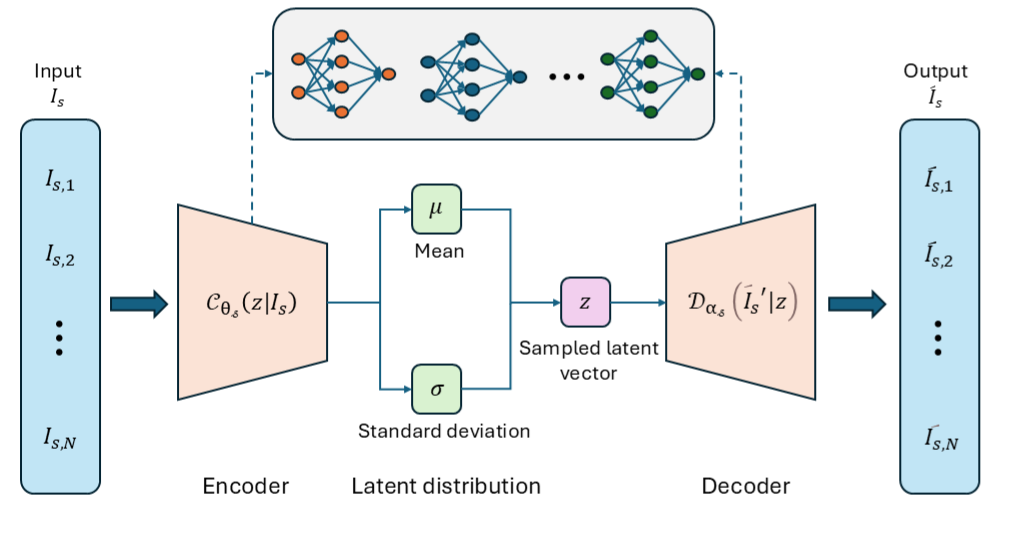}
    \vspace{-5mm}
    \caption{An illustration of the VAE structure for the image reconstruction with SCM selection.}
    \label{vae}
    \vspace{-6mm}
\end{figure}

In order to realize the dynamic SCM selection and resource allocation, we first train the SCMs so that they can reconstruct images in good quality. Different CNN architectures, which are tailored for image-related tasks, are applied as the semantic models. These classic CNN architectures are widely recognized as effective for processing image tasks, making them suitable for our scenario. We intend to apply the VAE structure as in Fig.~\ref{vae} for the image reconstruction, where different SCMs are chosen as the encoder, each with a corresponding decoder at the receiver. Particularly, the sum of KL Divergence Loss, structural similarity index (SSIM), and mean squared error (MSE) is defined as the loss function for training the VAE, which ensures that the reconstructed image closely matches the original while enforcing the latent variable distribution to approximate a standard normal distribution. We first denote MSE as
\begin{equation}
    \text{MSE}\ =\ \frac{1}{N}\sum_{i=1}^{N}\left(v_i-{\hat{v}}_i\right)^2,
\end{equation}
where $N$ is the number of pixels in the image; $v_i$ and ${\hat{v}}_i$ are the pixel values of the original and reconstructed images, respectively. In addition, the KL divergence loss is defined as
\begin{equation}
    \text{KL}\ =-\frac{\beta}{2}\cdot\frac{1}{N}\sum_{i=1}^{N}\left(1+\log\left({\sigma_i}^2\right)-{\mu_i}^2-{\sigma_i}^2\right)\ ,
\end{equation}
where $\mu_i$ and  $\log\left({\sigma_i}^2\right)$ are the mean and log-variance of the latent distribution; $\beta$ is a weighting factor for the KL loss. Furthermore, SSIM can be expressed as
\begin{equation}
    \text{SSIM}\ =\ \frac{(2\mu_v\mu_{\hat{v}}+C_1)(2\sigma_{v\hat{v}}+C_2)}{({\mu_v}^2{\mu_{\hat{v}}}^2+C_1)({\sigma_v}^2{\sigma_{\hat{v}}}^2+C_2)},
\end{equation}
where $\mu_v$ and $\mu_{\hat{v}}$ are the mean intensities of $v$ and $\hat{v}$; ${\sigma_v}^2$ and ${\sigma_{\hat{v}}}^2$ are the variances of $v$ and $\hat{v}$; $\sigma_{v\hat{v}}$ is the covariance between $v$ and $\hat{v}$; $C_1$ and $C_2$ are constants to stabilize the division. The total loss is subsequently modeled as
\begin{equation}\label{eq:vae_loss_function}
    \text{Loss}=\alpha\cdot \text{MSE}+\left(1-\alpha\right)\cdot\left(1-\text{SSIM}
    \right)+\text{KL}.
\end{equation}

This combined loss function balances pixel-wise accuracy, structural similarity, and latent space regularization, ensuring high-precision image reconstruction. After training, each model can encode images and decode them while also recording its computation energy consumption and latency during encoding.

\subsection{PPO Agent for SCM Selection and Resource Allocation}

In a real-world multi-user image transmission scenario, each user selects one of the trained SCM models. At this stage, the system is already aware of each model’s energy consumption and latency. The optimization goal is to maximize the RDE between the original and reconstructed images while satisfying constraints on total energy consumption, latency, and bandwidth. To dynamically select the optimal SCM model, allocate power, and bandwidth, we construct a customized RL environment in OpenAI gym~\cite{10333793}, which is specifically designed in the following:
\begin{itemize}
    \item \textit{State}: The state of the system at time $t$ is denoted by $s_t=\{\mathcal{C}_t,\mathcal{P}_t,\mathcal{B}_t\}$, which includes user channel conditions $\mathcal{C}_t=\{{C}_{t,1},C_{t,2},\cdots,\ C_{t,M}\}$, remaining power $\mathcal{P}_t=\{P_{t,1},P_{t,2},\cdots,P_{t,M}\}$, and remaining bandwidth $\mathcal{B}_t=\{B_{t,1},B_{t,2},\cdots,B_{t,M}\}$.% and image complexity $\mathcal{O}_t=\{O_{t,1},O_{t,2},\cdots,O_{t,M}\}$. Particularly, a combined score involving entropy, edge density, and Discrete Cosine Transform (DCT) energy measurement is proposed to quantify the image complexity. Entropy measures randomness and texture richness, which can be mathematically modeled as
%$H=-\sum_{i=1}^{N}{p_i\log{p_i}}$,
%where $p_i$ is the probability of pixel intensity i. Higher entropy means higher complexity. Edge density captures structure and sharpness, which can be measured by Canny edge detection. A higher edge density refers to more complex images. Additionally, DCT energy detects high-frequency details, which handles the differentiation of blurred and sharp images well.

\item \textit{Action}: Each user is asked to select a suitable SCM model for image transmission with intelligent power and bandwidth allocation. Particularly, for each SCM, the model name, power consumption, duration, and model quality are recorded, where the model quality is modeled by the Loss function when training the SCM. We denote $A_{m,t}$ as the set for the action space of selecting SCM and resource allocation for user $m$ at time $t$, which involves discrete parameters for SCM selection and continuous ones for power and bandwidth allocation.

\item \textit{Reward}: To quantify the action's performance, the system's reward is designed as the RDE, while penalties are applied for violating energy and latency constraints. Specifically, the reward function is defined as
\begin{multline}
    R =  \frac{\sum_{m=1}^M r_m}{\lambda\sum_{m=1}^M D_m+\varepsilon}-\gamma_1\cdot \biggl(\sum_{m=1}^M (p_{m,c}+p_{m,t})-P_{max}\biggr)\\-\gamma_2\cdot\biggl(\sum_{m=1}^M (\tau_{m,c}+\tau_{m,t}-\tau_{m,max})\biggr).
\end{multline}
\end{itemize}

For this environment, we use PPO to learn an optimal policy, enabling dynamic resource allocation and model selection. PPO is an RL algorithm based on policy gradients, designed to stabilize the training process by limiting the extent of policy updates. The core idea behind PPO is the use of a clipped policy update to prevent overly large changes in the policy, thereby improving both stability and efficiency in training.

PPO employs two NNs, a policy network $\pi_{\theta}$ which is responsible for selecting actions, and a value network $V_{\varphi}$ estimating the value of a given state.
Then, PPO calculates the advantage function using generalized advantage estimation (GAE), which evaluates the relative benefit of taking a specific action and can be numerically expressed as
\begin{equation}
     A(s_t, a_t) = G(s_t) - V_\varphi(s_t),
     \label{adv}
\end{equation}
where $G(s_t)$ represents the discounted return from state $s_t$ for each timestep.
To ensure training stability, PPO introduces a clipped surrogate objective to constrain policy updates:
\begin{equation}
        L_{\text{policy}}(\theta) = -\min\biggl(\epsilon_t A(s_t, a_t), \text{clip}(\epsilon_t, 1-\varepsilon, 1+\varepsilon) A(s_t, a_t)\biggr),
        \label{loss_policy}
\end{equation}
where $ \epsilon_t= \frac{\pi_\theta(a_t | s_t)}{\pi_{\theta_{\text{old}}}(a_t | s_t)}$ is the probability ratio between the new and old policy.
The value network is trained using a mean squared error loss function, which can be represented as
\begin{equation}
     L_{\text{value}}(\varphi) = \left(V_\varphi(s_t) - G(s_t)\right)^2.
     \label{loss_value}
\end{equation}
By combining \eqref{loss_policy} and \eqref{loss_value}, the total loss function can be expressed as
\begin{equation}
    L_{\text{total}}(\theta, \varphi) = L_{\text{policy}}(\theta) + c \cdot L_{\text{value}}(\varphi),
    \label{loss_total}
\end{equation}
where $c$ is the weight for value loss. %Then, the parameters $\theta$ and $\varphi$ can be updated according to the following rules:
%\begin{equation}
%    \theta = \theta - \alpha \cdot \nabla_\theta L_{\text{total}}(\theta, \varphi),
%    \label{theta_update}
%\end{equation}
%\begin{equation}
%    \varphi = \varphi - \alpha \cdot \nabla_\varphi L_{\text{total}}(\theta, \varphi).
%    \label{varphi_update}
%\end{equation}

\begin{algorithm}
\caption{PPO Algorithm for SCM Selection and Resource Allocation}
\label{alg:ppo_scm}
\begin{algorithmic}[1]
\State Initialize policy network $\pi_\theta$ and value network $V_\varphi$ with parameters $\theta$ and $\varphi$, respectively.
\State Initialize environment and PPO hyperparameters: learning rate $\alpha$; discount factor $\gamma$; GAE parameter $\lambda$; clipping range $\varepsilon$; number of updates per epoch $K$; number of trajectories per epoch $N$; maximum steps per trajectory $T$.
\For{epoch $= 1$ to \texttt{num\_epochs}}
    \For{episode $= 1$ to $N$}
        \State Initialize state $s_0$.
        \For{$t = 0$ to $T-1$}
            \State Select action $a_t$ according to current policy $\pi_\theta(a_t|s_t)$;
            \State Execute action $a_t$, observe reward $r_t$ and next state $s_{t+1}$;
            \State Store $(s_t, a_t, r_t, s_{t+1})$ in the experience buffer.
        \EndFor
    \EndFor
    \For{each trajectory in experience buffer}
        \State Compute advantage function $A(s_t, a_t)$ for each timestep using \eqref{adv}.
    \EndFor
    \For{$k = 1$ to $K$}
        \State Sample a random mini-batch %$\biggl(s_t, a_t, r_t, s_{t+1}, G(s_t), A(s_t, a_t)\biggr)$ 
        from the experience buffer;
        %\State Compute policy loss $L_{\text{policy}}(\theta)$ using \eqref{loss_policy};
        %\State Compute value loss $L_{\text{value}}(\varphi)$ using \eqref{loss_value};
        \State Compute total loss $L_{\text{total}}(\theta, \varphi)$ using \eqref{loss_total};
        \State Update $\theta$ and $\varphi$ using gradient descent.% based on \eqref{theta_update} and \eqref{varphi_update}.
    \EndFor
    \State Clear the experience buffer.
\EndFor
\end{algorithmic}
\end{algorithm}

By continuously updating the policy network and the value network, agents can select actions with higher rewards. The PPO-based joint SCM selection and resource allocation
algorithm is summarized in Algorithm~\ref{alg:ppo_scm}.

\section{Simulation and Results}
\label{sec:simulation and results}
In this part, numerical examples are demonstrated to verify the effectiveness of our system model and algorithm. We first train the VAE for image reconstruction with different encoder models and corresponding decoders. The model was trained with a combined loss function \eqref{eq:vae_loss_function} that includes MSE, SSIM, and KL divergence to balance reconstruction quality and latent space regularization. The trade-off parameter $\alpha$ in the TL function is set as $\alpha=0.8$. The training process utilized the Adam optimizer with a learning rate of $5^{-4}$ and weight decay of ${10}^{-5}$. Early stopping was implemented with a patience of $10$ epochs and a minimum improvement delta of $0.001$ to prevent overfitting. Each model was trained for a maximum of 200 epochs with a batch size of 128, and gradient clipping was applied with a maximum norm of 5.0 to ensure stable training. 

Particularly, we choose LeNet, ResNet, MobileNet, and DPN-26 models for the image reconstruction task. After training, we obtain the power consumption, duration, and the training objective function value, which are shown in the Table~\ref{table1}~\footnote{The models' power consumption and interference time data is captured from study~\cite{sanchez2025green} for testing of GPU.}.
\begin{table}[h]
\vspace{-4mm}
    \centering
    \caption{Simulation Parameters}
    \renewcommand{\arraystretch}{1.1} % Adjust row height slightly
    \small % Reduce font size to fit in one column
    \resizebox{\columnwidth}{!}{ % Resize table to fit within one column
        \begin{tabular}{
            >{\centering\arraybackslash}p{2cm}
            >{\centering\arraybackslash}p{2cm}
            >{\centering\arraybackslash}p{2.2cm}
            >{\centering\arraybackslash}p{2cm}
        }
            \toprule
            \textbf{Model} & \textbf{Power consumption (W)} & \textbf{Inference time on $5\times 10^4$ samples (s)} & \textbf{Loss (eq. \eqref{eq:vae_loss_function})} \\ 
            \midrule
            LeNet & 120 & 5.1 & 24.65 \\ 
            \midrule
            ResNet & 270 & 5.3 & 14.00 \\ 
            \midrule
            MobileNet & 170 & 5.3 & 24.93 \\ 
            \midrule
            DPN-26 & 305 & 9.5 & 10.76 \\ 
            \bottomrule
        \end{tabular}
    }
    \label{table1}
    \vspace{-3mm}
\end{table}

\begin{figure}[t]
    \centering    \includegraphics[width=0.8\linewidth]{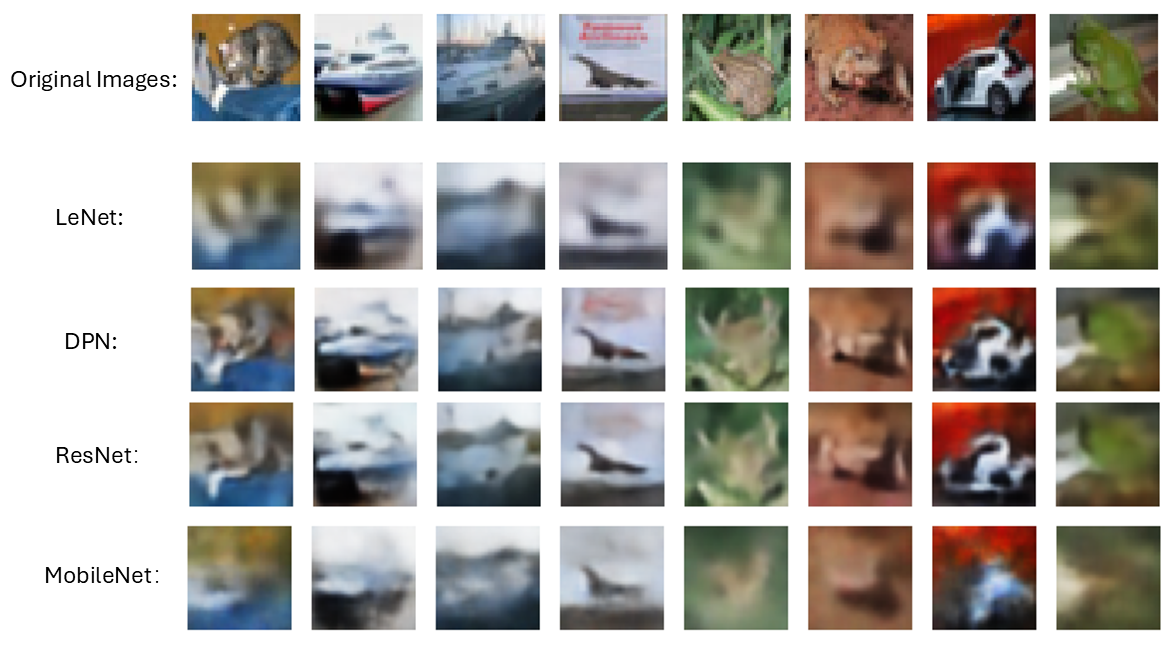}
    \vspace{-3mm}
    \caption{The reconstructed images with different CNN models.}
    \label{models}
    \vspace{-7mm}
\end{figure}

In parallel, Fig.~\ref{models} demonstrates testing examples of the image reconstruction under different trained SCMs. Models such as LeNet and MobileNet exhibit limited reconstruction quality, albeit with relatively low power consumption and computational latency. In contrast, DPN-26 achieves better reconstruction quality, albeit at the cost of higher power consumption and extended computational duration. To accommodate diverse user requirements and image characteristics, users should dynamically select appropriate encoder-decoder pairs for image reconstruction. Additionally, adaptive allocation of power and bandwidth is essential to ensure optimal QoS.

\begin{figure*}[t]
\centering
\begin{minipage}{.32\textwidth}{
\includegraphics[width=2.2in,height=1.7in]{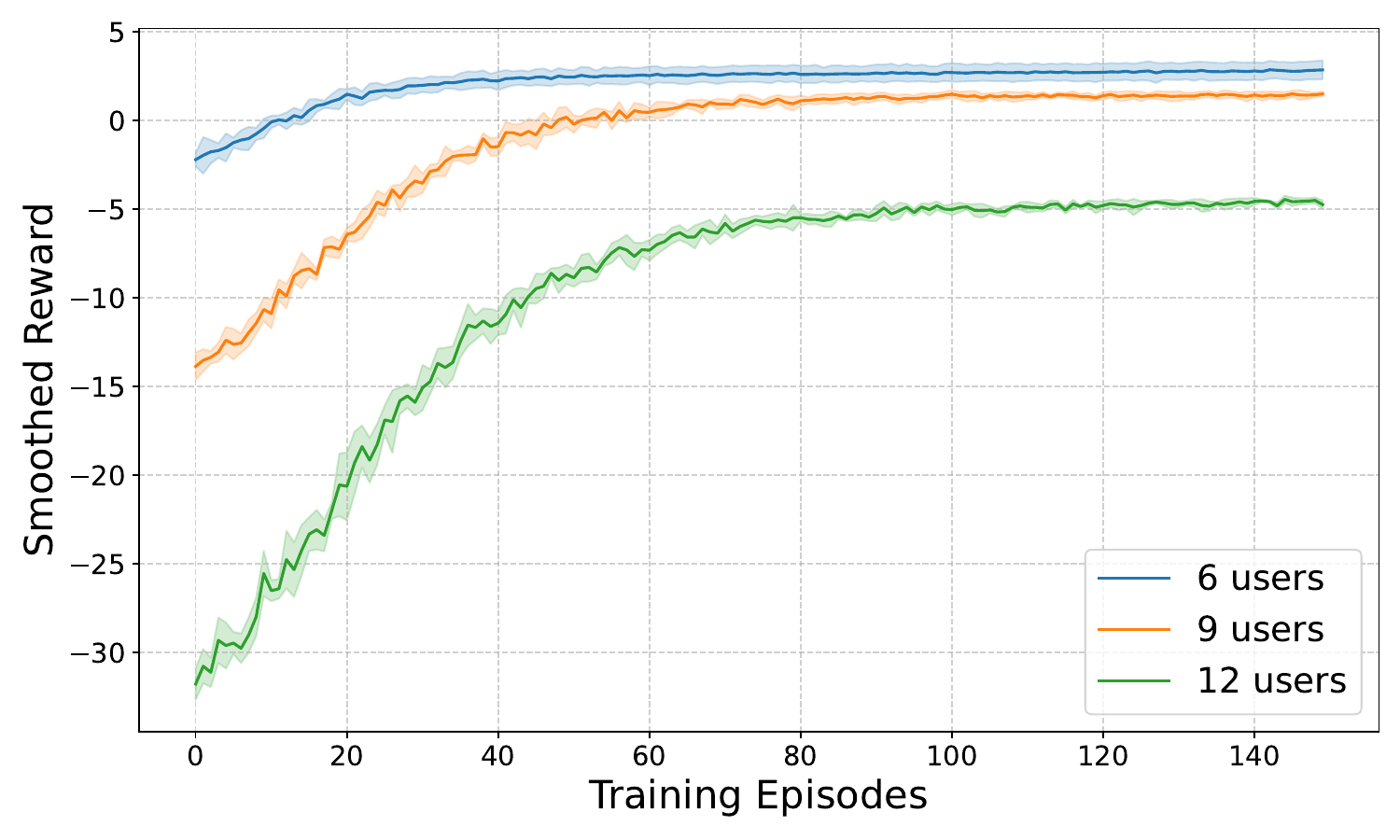}}\vspace{-4mm}
    \caption{Convergence behaviour of the proposed algorithm with different number of users.}
    \label{convergence}
\end{minipage}
\begin{minipage}{.32\textwidth}
{\includegraphics[width=2.2in,height=1.7in]{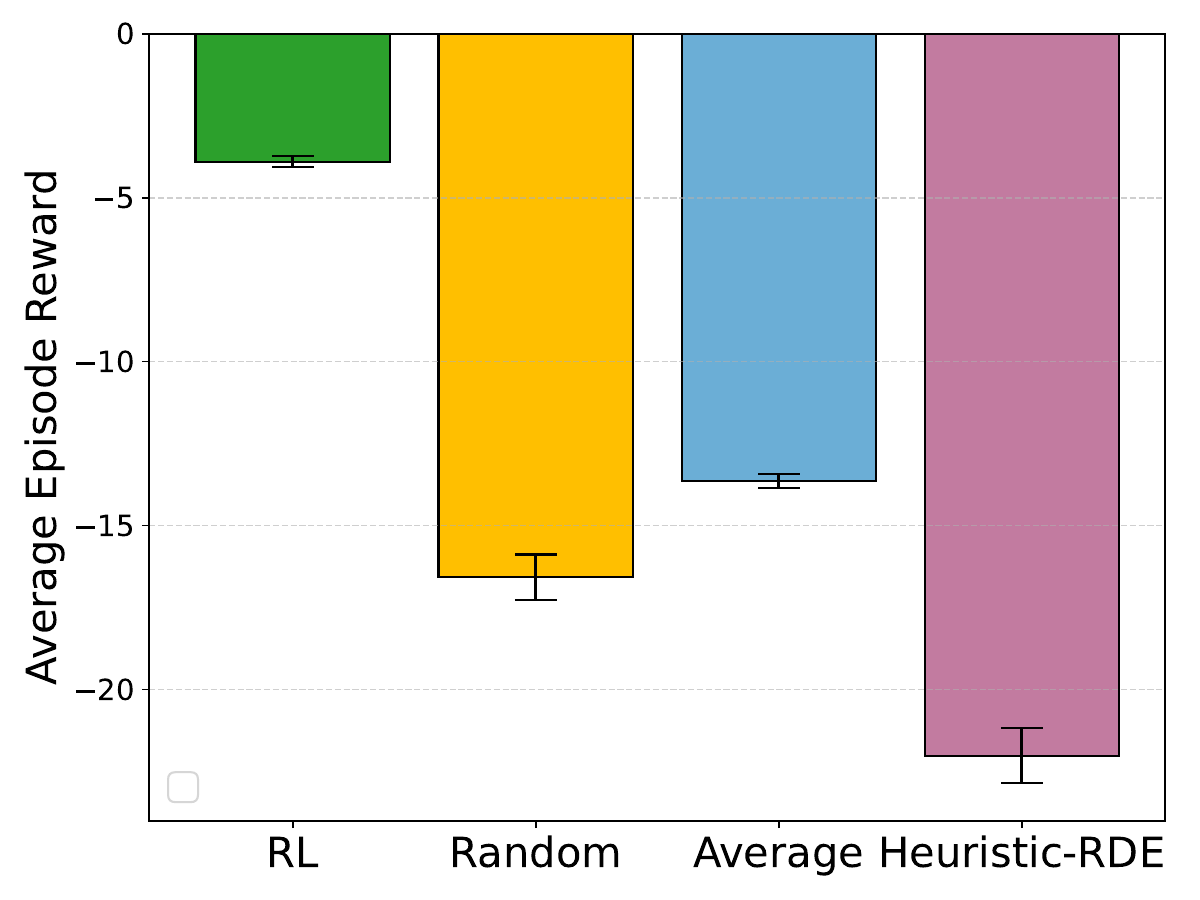}}\vspace{-4mm}
    \caption{Comparison of average episode reward and standard deviation across different strategies.}
    \label{compare1}
\end{minipage}
\begin{minipage}{.32\textwidth}
{\includegraphics[width=2.2in,height=1.7in]{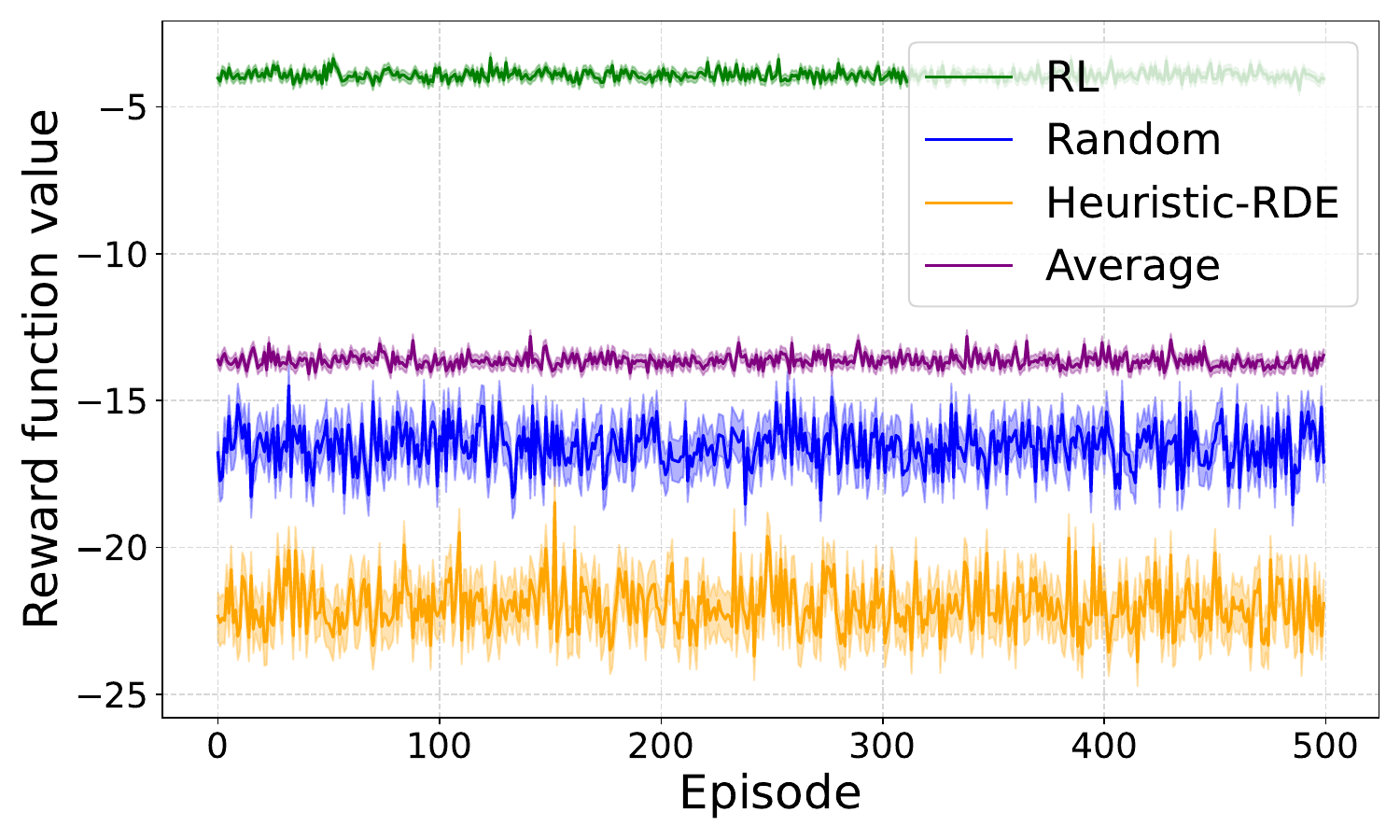}}\vspace{-4mm}
%\vspace{-3mm}
    \caption{Reward function value over eposides across different strategies.}
    \label{compare2}
\end{minipage}
\vspace{-5mm}
\end{figure*}
We then evaluate our proposed PPO algorithm for the SCM selection and resource allocation. Specifically, we assume an environment simulating under a different number of users, with a total system bandwidth of 30 MHz and a power budget of 3000 mW. Each user selects a pre-trained VAE model from a list of four (LeNet, ResNet, MobileNet, DPN), each associated with specific power consumption and latency parameters as given in Table.~\ref{table1}. Rayleigh fading channels are used with a standard deviation of 0.2. Since the distortion values based on MSE and SSIM are typically small, a large $\lambda$ is chosen to ensure that the distortion has a comparable impact on the reward scale as the rate term. Additionally, $\varepsilon$ is a small positive constant added to the denominator to ensure numerical stability and prevent division by zero. Therefore, the reward is set based on the RDE metric with $\lambda = 10^4$ and $\varepsilon=10^{-5}$, penalized for latency exceeding $\tau_{m,max} = 12$ s and for power deviation from the total budget $P_{max}$, weighted by $\gamma_1=0.3$ and $\gamma_2 = 0.2$. PPO training is configured with a learning rate of $3^{-5}$, 2048 steps per rollout, batch size of 64, 15 epochs per update, and $\gamma=0.995$, across 150 epochs with 2000 timesteps each. Fig. \ref{convergence} demonstrates the convergence behavior of the reward function when the number of users is set as 6, 9, and 12, respectively, revealing the effectiveness of the proposed PPO algorithm. The reward curve is smoothed with interval confidence over five random experiments. 

Fig.~\ref{compare1} illustrates the average episode rewards over 400 test episodes with 12 users per episode achieved by four different strategies: (a) our proposed PPO-based RL; (b) random:  selects the SCM model index and allocates resources randomly; (c) average: assigning the same SCM model to all users and equally dividing bandwidth and power among them; and (d) heuristic-RDE: greedily selects the SCM model for each user based on the model’s individual RDE, while bandwidth and power are uniformly allocated. Among the methods, the proposed RL strategy demonstrates the best overall performance, achieving a mean reward of $-3.92$ with a small standard deviation of $\pm 0.17$, indicating both high performance and consistency. In contrast, heuristic-RDE, which greedily selects the SCM model per user based on a simplified RDE metric, performs the worst with a mean reward of $-22.06$ and a standard deviation of $\pm 0.78$, highlighting its inability to effectively balance system-level trade-offs. The average and random strategies show intermediate performance, scoring $-13.64\pm 0.21$ and $-16.53\pm 0.73$, respectively, but still significantly underperform compared to the learned RL policy. These results clearly demonstrate the advantage of reinforcement learning in capturing complex environment dynamics and optimizing long-term objectives over naive or greedy baselines.

Fig. \ref{compare2} further visualizes the episode-wise reward trajectories for each method during the inference stage. The lines represent the average reward per episode, while the shaded regions indicate the standard deviation, capturing the variability of each strategy’s performance at each episode. The RL method consistently achieves the best rewards, demonstrating strong learning-based adaptation. The heuristic-RDE approach, though optimal per user, often violates global resource constraints and incurs penalties, making it even less effective overall than the random strategy, which better preserves resource feasibility. This plot effectively emphasizes the robustness and superiority of our proposed method in complex resource allocation scenarios.

\section{Discussion}
% In this section, we first explore the proposed method from an implementation perspective and then analyzes its generalization capability, computational complexity, and scalability when applied in practical semantic communication systems.

% In the implementation workflow, the system components, namely, the users, the RAN, and the edge server, act in distinct yet collaborative roles within the network architecture in the image transmission process. User devices are responsible for capturing images and conducting lightweight semantic encoding. 
% 
% They transmit the extracted semantic features, along with information such as channel state information (CSI), latency feedback, and energy consumption metrics, to the RAN. The RAN then functions as an intermediary, forwarding data to the edge server while also aggregating real-time network statistics, including channel quality and resource usage. At the edge, a DRL agent resides, which learns a policy to jointly optimize SCM selection and resource allocation across multiple users, aiming to maximize long-term system performance considering both the image reconstruction quality and transmission efficiency.

\textit{\textbf{Generalization Ability:}} The strength of this DRL-based semantic communication framework lies in its potential to generalize across diverse user behaviors, dynamic channel conditions, and varying semantic content. This is largely attributed to its feature-driven design and task-oriented optimization objective. Nevertheless, its generalization may degrade when deployed in scenarios that differ significantly from those seen during training—such as new semantic domains, user distributions, or network configurations. This is usually termed as the \textit{sim2real} problem of DRL. It is important to train the agent with diverse, randomized conditions to enhance its adaptability, and the policy transfer techniques such as transfer learning, meta learning or incremental learning can be used. Meanwhile, these methods should be integrated with the in-network ML/RL operations for realistic implementation~\cite{li2022rlops}.

\textit{\textbf{Complexity:}}
From a complexity standpoint, although DRL training, especially with algorithms like PPO, can be computationally intensive due to the large and continuous state-action space, the inference phase is highly efficient. Once trained, the policy network supports low-latency decision-making, which is suitable for real-time applications. For instance, the DPN-26 model takes 40.8 seconds to train for one epoch, yet inference on 50,000 samples completes in only 9.5 seconds \cite{sanchez2025green}, demonstrating its practicality in latency-sensitive environments.

\textit{\textbf{Scalability:}}
Scalability is another key consideration in practical deployment. While the framework inherently supports multi-user joint optimization, its action space expands with the number of users and available SCM options, which may slow training or require more sophisticated policy architectures. Techniques such as parameter sharing, hierarchical control, and decentralized multi-agent RL can help alleviate these limitations, ensuring the framework remains effective as system size increases. Taken together, the proposed method offers a balanced trade-off between semantic accuracy, computational efficiency, and deployment flexibility, making it a viable candidate for scalable semantic communication systems.

\textit{\textbf{From Semantic Compression Model to Reconstruction Model:}}
In this work, we primarily focus on the selection of the semantic model for compression. Leveraging the symmetric property of semantic model pairs, the semantic decoder can be aligned with the chosen compression model to ensure robust reconstruction. Additionally, the semantic decoder can be optimized independently on the receiver side, without considering radio resource constraints, to achieve a better trade-off between computational complexity, reconstruction quality, and latency.
\vspace{-5mm}
\section{Conclusions}
\label{sec:conclusion}
In this paper, we investigate a SemCom system comprising multiple users, a RAN, and an edge server under an image transmission setting. To jointly optimize the selection of SCMs and resource allocation strategies, we propose a DRL approach. A novel metric, RDE, is introduced to capture the trade-off between image reconstruction quality and transmission efficiency. We adopt a PPO-based reinforcement learning framework to optimize the RDE at the system level. Extensive simulations demonstrate the superiority of the proposed method over several baseline strategies. Additionally, we provide a comprehensive discussion on generalization ability, computational complexity, scalability, and insights into the design of semantic communication models.
\vspace{-1mm}
\section*{Acknowledgment}
This work is supported by the 6G-GOALS project under the 6G SNS-JU Horizon program, n.101139232.
\vspace{-1mm}
\bibliographystyle{IEEEtran}  % 或其他样式
\bibliography{references}     % 假设你的 .bib 文件名叫 references.bib

% \section{biographies}
% \begin{IEEEbiographynophoto}
%   XX
% \end{IEEEbiographynophoto}
\end{document}